\documentclass[manuscript,nonacm]{acmart}

\AtBeginDocument{%
  \providecommand\BibTeX{{%
    \normalfont B\kern-0.5em{\scshape i\kern-0.25em b}\kern-0.8em\TeX}}}

\begin{document}

\title{Youth WellTech: A Global Remote Co-Design Sprint for Youth Mental Health Technology}

\author{Kenji Phang}
\email{chang.phang@student.manchester.ac.uk}
\affiliation{%
  \institution{University of Manchester}
  \city{Manchester}
  \country{United Kingdom}
}

\author{Siddharth Saarathi Pradhan}
\email{sidspedx@gmail.com}
\affiliation{%
  \institution{Guru Nanak institutions Technical Campus}
  \city{Hyderabad, Telangana}
  \country{India}
}

\author{Chino Ikwuegbu}
\email{chinoikwuegbu@gmail.com}
\affiliation{%
  \institution{University of Port Harcourt}
  \city{Choba, Rivers}
  \country{Nigeria}
}

\author{Gonzalo Ramos}
\orcid{0000-0003-4198-5021}
  \email{goramos@microsoft.com}
\author{Denae Ford}
  \email{denae@microsoft.com}
\affiliation{%
  \institution{Microsoft Research}
  \city{Redmond, WA}
  \country{USA}}
  
\author{Ebele Okoli}
\affiliation{
\institution{Microsoft}
\city{Atlanta, GA}
\country{USA}
}
\email{ebeleokoli@microsoft.com}

\author{Salman Muin Kayser Chishti}
\email{salman.chishti@microsoft.com}
\orcid{0009-0007-9949-3560}
\affiliation{%
  \institution{Microsoft}
  \city{Talinn}
  \country{Estonia}
}

\author{Jina Suh}
\email{jinsuh@microsoft.com}
\affiliation{%
  \institution{Microsoft Research}
  \city{Redmond}
  \country{USA}
}

\renewcommand{\shortauthors}{Phang, et al.}

\newcommand{\todo}[1]{\textcolor{red}{[TODO: #1]}}
\newcommand{\jina}[1]{\textcolor{purple}{[Jina: #1]}}
\newcommand{\salman}[1]{\textcolor{orange}{Salman: #1}}
\newcommand{\yousra}[1]{\textcolor{cerulean}{Yousra: #1}}
\newcommand{\kenji}[1]{\textcolor{blue}{Kenji: #1}}

\newcommand{\squeeze}[1]{\textls[-10]{#1}}
\newcommand{\squeezemore}[1]{\textls[-20]{#1}}
\newcommand{\squeezewaymore}[1]{\textls[-40]{#1}}

\newcommand{\xhdr}[1]{\textit{#1}\hspace{0.5em}}

\hyphenpenalty=10000
\brokenpenalty=10000
\sloppy
\raggedbottom

\begin{abstract}
Mental health is a pressing concern in today's digital age, particularly among youth who are deeply intertwined with technology. Despite the influx of technology solutions addressing mental health issues, youth often remain sidelined during the design process. While co-design methods have been employed to improve participation by youth, many such initiatives are limited to design activities and lack training for youth to research and develop solutions for themselves. In this case study, we detail our 8-week remote, collaborative research initiative called Youth WellTech, designed to facilitate remote co-design sprints aimed at equipping youth with the tools and knowledge to envision and design tech futures for their own communities. We pilot this initiative with 12 student technology evangelists across 8 countries globally to foster the sharing of mental health challenges and diverse perspectives. We highlight insights from our experiences running this global program remotely, its structure, and recommendations for co-research.
\end{abstract}

\maketitle

\section{Introduction and Background}

The mental well-being of today's youth is a pressing concern, underscored by alarming statistics. 
According to the World Health Organization~\cite{who}, approximately 20\% of the world's children have a mental well-being condition, suicide is the second leading cause of death amongst 15-29 year-olds~\cite{Rosenberg1987The}, and one in seven 10-19 year-olds globally experience mental well-being concerns, many of which go unrecognized and untreated.
Furthermore, the COVID-19 pandemic has exacerbated these concerns. A 2021 meta-analysis of 29 studies~\cite{Hawes2021Increases} found symptoms of depression and anxiety to be "significantly higher" following the onset of the pandemic.
Many scholars have attributed the prevalence of youth mental well-being issues to the growing significance of technology~\cite{stergaard2017TakingFA,Kappos2007The,mahel2015TheIO} in their lives that may amplify social comparison~\cite{Vannucci2019Social,Liu2021WhyAY,Fardouly2017The} and technology addiction~\cite{Foerster2015Problematic} through designs that capture youth's attention and promote endless engagement~\cite{meyerson2012youtube,covington2016deep}.
On the other hand, others argue that technology can also provide many benefits to youth~\cite{Valaitis2005Computers, Abuwalla2017Long-term} such as providing them with the ability to work remotely with others around the world~\cite{Dwyer2017Smartphone}, offering educational and career opportunities~\cite{Suyatna2018Sociopreneurship}, and giving them a platform to raise their voice~\cite{Mohamad2018Youth,Marchi2018Social}. 
Growing up immersed in technology, today's youth offer invaluable firsthand experiences, highlighting the importance of their involvement in research, design, and development of technologies.

Despite the potential of youths' insights on the creation of new solutions, youths are often sidelined in technology design and implementation.
Co-design, a collaborative approach to problem-solving, has emerged as a promising avenue to bridge the gap in addressing youth mental well-being issues~\cite{Garner_2021}. It fosters diverse discussion between groups of people which may otherwise be overlooked due to bias of an independent researcher~\cite{Trischler2018Team, agbo2021co}, and through actively involving all stakeholders, co-design aims to create more relevant and effective solutions. 
Historically, co-designing with youths has been shown to be a successful approach to tackling the complex intricacies in society that are challenging to cover through traditional methods~\cite{OBrien2020ASR}. 
Furthermore, it also inherently benefits the community, especially in enhancing mental well-being~\cite{OBrien2020ASR}.
Co-designing with youth has shown additional benefits in developing novel design features and gaining insights into adolescent attitudes towards various scientific concepts, such as being curious through proposing prototypes and asking questions, whilst not being well versed in the area~\cite{Bonsignore2016Traversing}. 

Our case study explores two under-addressed areas in co-design.
The first area is focusing on global participation. 
Most co-design programs are localized, focusing on a particular community, region, or culture~\cite{burkett2012introduction, wake2013developing, locock2014testing}. 
However, in the case of mental well-being technology, it is important to work in an integrated manner with people cross-culturally because technology solutions are often motivated by their potential to increase accessibility and scale globally and because addressing mental well-being requires understanding individual and cross-cultural differences~\cite{James2002Cultural, Ochnik2021ACO}. 
Yet, global co-design brings complexity, from cultural communication barriers~\cite{Sanchez-Burks2003Conversing} to challenges in remote interactions, like misreading online social cues~\cite{Wei2012A}. 
Even for experienced speakers and the youth, existing social barriers, such as the fear of judgment or reluctance to voice disagreements, can be magnified in remote settings~\cite{Slater2006AnES,Hilliard2020StudentsEO}. 
Our study specifically targets a global cohort, with a goal to explore the challenges and merits of remote co-design sessions.

The second area is maximizing the level of involvement by youth to include the development of technological designs and the analysis of outputs that arise from co-design activities. 
Typically, co-design sees researchers translating outputs into the development of new solutions while participants give iterative feedback on problems, ideas, and developed technology prototypes via workshop activities~\cite{Education_2022}.
Prior research has raised several limitations of traditional co-design approaches. For example, tokenism, where participants are merely symbolic without genuine involvement, can dilute the essence of co-design~\cite{Molle038339}. Power dynamics can persist, making participants, especially youths, hesitant to voice their opinions~\cite{uchidiuno2023s,farr2018power,sbaiti2021whose}, or insufficient training can leave participants feeling ill-equipped~\cite{Molle038339,peters2018participation}.
Previous work has explored engaging youth as co-researchers to raise the level of their involvement~\cite{kim2016youth, genuis2015partnering, clark2022moving}.
In addition, with educational programming coupled with the use of modern prototyping tools that lower barriers of entry~\cite{cross2013visual,kross2022five,ribeiro2012survey}, it is more feasible now than ever to empower youth to be more involved in the design and implementation process.  
Our study targets technologically proficient youth~\cite{castellacci2011closing}, such that their existing technical understanding can accelerate them to conceive innovative solutions using technology. Our approach champions ``Youth-Driven Technology Design'', empowering the youth with more ownership, all the way from ideation to synthesis. It embodies a design philosophy driven by the youth, for the youth, and through the youth, centering their voices deep within the design process.

Building on these two identified gaps, we conducted a remote, 8-week collaborative research program with global youth to co-design mental well-being technologies by equipping them with Human-Centered Design (HCD) skills that can augment their existing technical skills.
Our program also empowers them to be co-researchers beyond design participants by helping them reflect on their experiences, analyze the generated artifacts (e.g., transcripts, reflection surveys), and refine co-design and co-research methodologies with youths, culminating in authoring this paper.
Through this collaborative experience, we aimed to answer the following research question: How can we effectively conduct co-design with a remote and diverse cohort of youths, ensuring meaningful participation, addressing cultural nuances, and empowering them to research and build their own solutions?
In this paper, we present a comprehensive case study of our co-design program, detailing our processes, encountered challenges, and the invaluable lessons learned. Our co-design research not only contributes a unique perspective by actively involving youths in addressing mental well-being challenges but also offers a diverse, remote collaboration with students from varied backgrounds. Furthermore, our findings provide actionable insights and recommendations for refining co-design research methodologies in the future, setting a precedent for how such collaborative activities can be effectively conducted moving forward.

\section{Youth WellTech} 

We developed the Youth WellTech as a pilot program consisting of youth participants who provide expertise and guidance for technology design for the well-being of the youth population. 
The members of the program are empowered with human-centered design practices and research mentors to investigate youth well-being challenges and to co-design future technology solutions to address them.
The primary goals of the program are as follows: (1) Recruit globally to generate diverse perspectives in problem formulation and design ideation, (2) Co-design by youth for youth through youth, (3) Empower youth with HCD skills to enable them to build solutions themselves, and (4) Empower youth to reflect, synthesize, and revise the program.

The program structure draws from the human-centered design process~\cite{viswanathan2004community, cornwall1995participatory}, specifically the design sprint methodology~\cite{horowitz2009community}, created and popularized by Google Ventures\cite{googledesignsprint}. 
Our program structure closely emulates the design sprint process but with two modifications. Firstly, we tailor the design sprint process for co-design, where the design process is a shared process with end-users or stakeholders being active participants, to create a shared understanding of the problem, identify potential solutions, and develop and test prototypes. Secondly, we elevate the role of the end-users or stakeholders as co-researchers to democratize knowledge production~\cite{wallerstein2006using, jagosh2012uncovering}, recognizing their expertise and agency in the research and design process, and provide ownership over the research outcomes.

\begin{figure}[h]
\centering
\includegraphics[width=\textwidth]{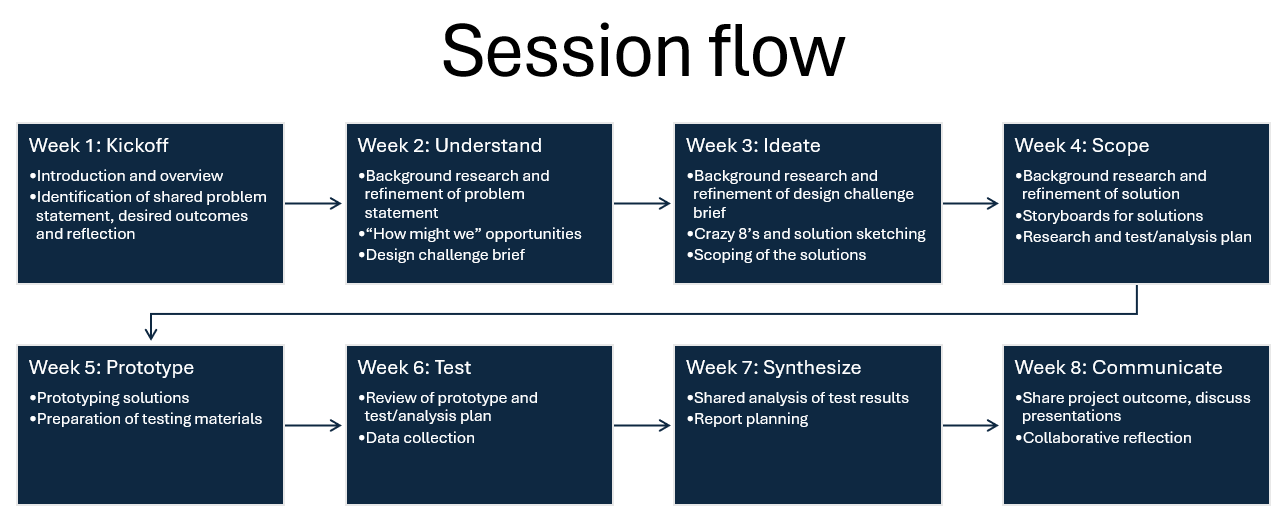}
 \caption{A graphic showing the flow of the sessions that we carried out in the Youth WellTech program, which was outlined to the participants in the first meeting and closely matched the execution.}
 \Description{A graphic showing the flow of the seessions that we carried out in the Youth WellTech program, which was outlined to the participants in the first meeting and closely matched the execution.}
\label{fig:session_flow}
\end{figure}

\subsection{Recruitment and participants}

We partnered with a student ambassador program of a large technology company to recruit globally diverse youth participants who were also at the forefront of developing and engaging with technology. 
We envisioned this partnership as a foundation for long-term engagement with youth in technology for collaborative designing and developing of mental well-being solutions.
The student ambassador program enrolled a global cohort of students from around the world with diverse backgrounds, cultures, and interests, united by a common goal and passion for technology, social impact, and enthusiasm to teach others.
Our call for participation included an interest form that asked for a brief statement on youth mental well-being interests, time availability, basic demographics (age and education), and prior experience in HCD or mental well-being-related work.
We received a total of 48 applications across 20 different countries. 
Three program organizers iteratively scored the applications on 3 dimensions: (1) how well they expressed their passion for MH in their statement, (2) the feasibility of the proposed approach in their statement, and (3) general time availability for sessions. 
Sampling from the ranked list of potential candidates with the highest total scores, we maximized the diversity of the group.

In total, 12 candidates consented to participate in the pilot program, with 6 self-reporting as female, 5 as male, and 1 non-binary/gender-diverse and spanning 8 countries (Algeria, Bangladesh, India, Kenya, Mexico, Nigeria, North Macedonia, Pakistan, UK). 
The average age of the 12 participants was 22.6 years old (min: 21, max: 27), and 10 out of 12 participants self-disclosed as having mental well-being lived experience.
9 participants were enrolled in an undergraduate program, 2 were in a postgraduate program, and 1 recently graduated from an undergraduate program.
8 participants self-reported as having some course-based familiarity with HCD, 4 had research experience with HCD, 3 had conducted human subjects research, and 1 had no prior knowledge of HCD.
In addition to 12 student participants, the pilot program included 5 organizers, where 4 were researchers in Human-Computer Interaction (HCI) and Informatics and 1 was a former student ambassador. 
At the conclusion of the pilot program, 5 student participants opted to conduct an analysis of the program output and co-author this paper.

Acknowledging that everyone participated in the program activities, we refer to the organizers and facilitators as ``mentors'', the youth participants as ``students'', and the mentors and students who participated in this writing as ``co-authors''. 
The study design and protocol were approved by the Microsoft Research Institutional Review Board (IRB). 

\subsection{Program overview} 
Our pilot program was conducted over 8 weeks with 2-hour sessions per week. 
Mentors set up a Microsoft Teams\footnote{https://www.microsoft.com/microsoft-teams} instance where meetings, calendars, chats, and documents could be shared.
Each session was conducted remotely via Microsoft Teams video conferencing software, and collaborative activities were facilitated primarily using FigJam\footnote{https://www.figma.com/figjam/}.
Figure~\ref{fig:session_flow} provides an overview of the activities covered in each week.

The first two weeks focused on problem formulation. 
In the first week, students were introduced to co-design and its essential elements, which included getting to know each other and an in-depth overview of the program's objectives. This phase established a comprehensive framework, underpinned by design principles such as HCD and the practical utilization of tools like FigJam. A key activity had students engaged in a thought-provoking exercise to envision both utopian and dystopian scenarios about technology's impact on mental well-being. Insights from this activity were captured onto note cards on a FigJam whiteboard and later used in framing and refining the problem statements.
In the following week, students reflected on their individual and shared objectives for the program, laying the foundation for a clear problem statement. Together with mentors, students reviewed prior research and were introduced to the ``How Might We'' (HMW) technique and Simon Sinek's Golden Circle framework~\cite{hennauxcan} to aid in brainstorming ideas for potential solutions to youth mental well-being challenges and sharing inspirations for their ideas. Students were then tasked with homework to conduct background research related to the potential solutions they had brainstormed.

The next two weeks focused on ideation and sketching. 
The third week centered on discussing and integrating the background research done from the previous week. Mentors and students discussed techniques to simplify complex problems into manageable components and devise solutions. The week's main activity involved the `4 Step Sketching' technique, a storytelling approach that helped students visualize potential solutions. Homework required students to sketch a solution, craft a comic-like user story depicting the problem and its resolution, and further explore their understanding of the problem space.
Week four was crucial as students began prototyping and defining solution scopes. They, along with mentors, reviewed the sketches from the previous week. Prototyping encompassed various stages, including paper prototypes, wireframes, and UI mockups, with an emphasis on leveraging the Figma tool for digital prototyping. Based on the favored solutions, students formed three teams focusing on (1) setting guardrails through filtering unhealthy social media content, (2) addressing social media addiction through gamification, and (3) engaging in simple and meaningful activities to enhance productivity.
With an understanding of the chosen directions, mentors and students proceeded to explore the intricacies of solution scoping and planning. 
Students within groups worked independently on elevating their initial sketches into detailed prototypes.

Weeks five to seven focused on prototyping, testing, and synthesizing learnings from their experience. 
In week five, students shared their prototypes and received feedback from the group. The mentors introduced usability testing to evaluate prototype effectiveness, focusing on methods such as surveys or interviews with think-aloud to extract user feedback. Each group was given a usability testing template to devise a test plan.
In week six, students were paired to engage in iterative testing sessions to scrutinize each other's prototypes, offering suggestions and constructive feedback. Students further iterated on their prototypes in groups. 
The penultimate week witnessed the students transform their conceptualized solutions into tangible outputs. Students opted for video presentations as the preferred format, a decision informed by the desire to enhance video pitch creation and communication skills. This choice was motivated by the objective of creating impactful, accessible, and engaging content. In addition to output selection, mentors and students delved into the essential techniques of crafting effective video pitches.

In the final week, teams showcased their videos to key stakeholders, including research leaders, organizers, and project managers, ensuring a comprehensive review of the students' work. Each presentation covered their design journey and showcased the solutions they had developed. The satisfaction expressed by key stakeholders underscored the success of the co-design approach and the impactful outcomes achieved throughout our 8-week endeavor.

\begin{figure}[t]
    \centering
    \includegraphics[width=0.8\linewidth]{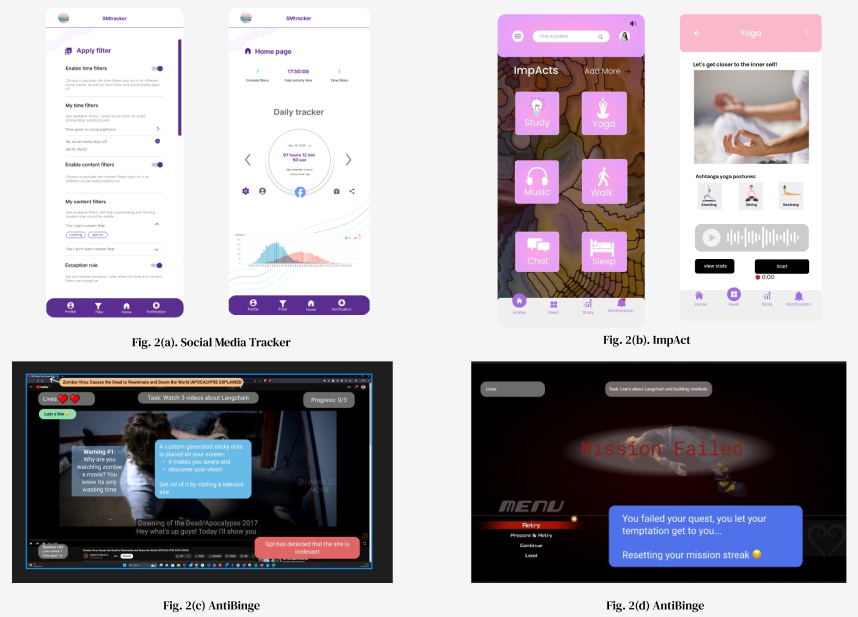}
    \caption{Prototypes generated as a result of the collaborative design process. (a) The Social Media Tracker (SMT) mobile app, designed to promote responsible and healthier social media usage through monitoring and regulation features.(b) The prototype  ImpAct, a mobile app promoting holistic mental wellness through a diverse catalog of activities and guidance (c) A visual representation of AntiBinge, displaying real-time notifications related to ongoing quests. (d) This is followed by another representation indicating that your quest or mission has not been successful.}
     \Description{Screenshot/glimpse of prototype designs.}
    \label{fig:prototypes}
\end{figure}

Three prototypes that the students generated are as follows:

\textbf{Social Media Tracker}: Social Media Tracker (SMT; Figure~\ref{fig:prototypes}a) is a mobile application designed to mitigate the excessive time spent on social media platforms.
SMT is designed to provide users with a strategic intervention that can be instrumental in fostering healthier social media usage patterns among youth.
It provides tools to monitor and regulate their social media activity through the incorporation of built-in time and content filters. Furthermore, the system incentivizes users through rewards based on their achievements in time-limited challenges.
By limiting digital dependency and promoting responsible usage, SMT aims to contribute positively to mental well-being. SMT empowers young individuals to track their daily social media consumption, utilizing time filters to curtail excess usage and providing notifications if necessary. Additionally, users can personalize their visible social media content with the aid of mental well-being tips integrated into the app.

\textbf{ImpAct}: ImpAct (Figure~\ref{fig:prototypes}b) is a mobile application engineered to propose a digital paradigm for prioritizing mental wellness. The fundamental aim of ImpAct is to actively foster individuals' investment in their mental well-being by diverting from their habitual routines and embarking on novel activities and experiences. In so doing, the application strives to advocate a holistic approach to mental wellness, a facet of health that frequently remains overlooked. The application furnishes a catalog of activities encompassing both physical (e.g., sports) and mental (e.g., yoga, reading) domains, many of which users may not have previously attempted. For example, for the choice of yoga, the app guides users with commencing the practice, understanding warm-up routines, mastering postures, and executing them correctly. This guidance is instrumental in ensuring the successful execution of tasks, reducing barriers to entry, and enhancing the overall user experience.

\textbf{AntiBinge}: AntiBinge (Figure~\ref{fig:prototypes}c-d) represents a gamified productivity application developed to address excessive consumption of detrimental YouTube content. The core concept of the app revolves around ``quests,'' which draw inspiration from traditional gaming while mirroring the real-life objectives and activities of the user. When a user selects a task from their to-do list, AntiBinge diligently monitors their YouTube usage, employing AI to analyze video metadata. As users remain aligned with their chosen objectives, they are rewarded with points as positive reinforcement. However, if the content deviates from the selected task, users lose lives, akin to the mechanics found in classic video games. When all lives are depleted, a custom-tailored motivational message is displayed, serving as a reminder of the user's original purpose for using the app. AntiBinge empowers young individuals to cultivate mindful content consumption, ensuring that their goals remain at the forefront while fostering resistance to the temptation of consuming unhealthy videos.

\subsection{Data collection and analysis}

Data for this paper was sourced through multiple avenues:
All sessions were recorded and transcribed using Teams' built-in features.
Chat histories from our shared Microsoft Teams channel and other chats were archived.
Session artifacts like sticky notes and sketches were saved in our shared FigJam.
Documents, including prototype specifications and relevant research papers, were stored in Teams.
After each session (Weeks 1-7), students filled out a weekly reflection survey. This survey asked about what activities they participated in, what went well or not, what they learned, what can be improved, and how. 
After the last week, the students were asked to complete a final reflection that asked for feedback on their shared project and the overall program, including how the program supported or failed to support their participation, how the program incorporated or failed to incorporate the voice of the youth in the design of technology for mental well-being, challenges faced or criticisms about the program, and benefits gained or praises about the program. 
Although reflections were encouraged, they were not mandatory, we were able to collect 50 weekly and 10 final reflections.

\squeezemore{Post-program, co-authors met for 8 weeks to analyze the collected data using reflexive thematic analysis~\cite{braun2023thematic} and descriptive quantitative analysis.
Transcripts and chats were reviewed by two co-authors, a different pair of co-authors evaluated the reflection data, and another two co-authors examined the session artifacts and final prototypes.
In weekly meetings, all co-authors, including mentors, collaborated to deduce common themes, informing co-design research implications.}

\section{Lessons Learned from Youth WellTech}
Overall, weekly feedback portrayed the program positively. Students praised its engaging structure, highlighting the program's educational focus on mental well-being, HCD techniques, and the effective use of online collaborative tools like Figma (e.g., ``Discussion and engagement activities were amazing... So much enthusiasm was there.''). The program fostered rich, cross-cultural conversations on youth mental well-being, with strong mentor support (e.g., ``It supported my participation through availability and the support from mentors and communication with fellow students.'') Moreover, the program inspired students to apply their learning externally. For instance, one student initiated their own blog post series with another highlighting ``The program's emphasis on engagement and commitment ... has spilled over into other areas of my life, helping me to develop a stronger work ethic and a greater sense of discipline.''

However, executing a global co-design program posed unique challenges. Social and hierarchical barriers in a digital setting sometimes heighten awkwardness. A participant voiced, ``Feels really awkward when no one is speaking, hard to judge the social cues'' This occasionally made it challenging to be open and candid, which is crucial for co-design. Technical issues and time zone differences added to the challenge. One participant shared, ``It was a bit difficult to work after the weekly sessions with the other team members as the time zones were different.'' Juggling academic and program responsibilities also stretched some students thin, leading to reduced active participation over time.
Addressing these challenges effectively is vital for genuine student involvement and program success. Below, We discuss our findings and offer lessons for future co-design efforts.

\subsection{Co-design activities should be fun and engaging.}
\squeeze{Students in our program, coming from various global locations, sometimes attended sessions after long days or late at night.  Given their potential fatigue, it is essential to structure sessions to be both intellectually stimulating and enjoyable.}

\xhdr{Potentially include games within co-design sessions.}
Including games within sessions serves a dual purpose: rejuvenating students and promoting camaraderie that overcomes virtual and cultural barriers. In our case, we noticed that students were really quiet in the early weeks. By the 3rd week, we introduced games like Skribble.io and CodeName, dedicating $\sim$25\% of our weekly meetings to them. While this may seem counterproductive, setting aside time for such bonding activities becomes a strategic investment, this led to positive feedback from students (e.g., ``Scribble has been fun'' ``breaking the barrier thing has been nice''), showcasing the tangible benefits of incorporating enjoyable components. 
In a long-term program like ours, spanning 8 weeks, including games prove beneficial.
However, the decision to incorporate games and the ideal balance between work and play depends on the program's length and objectives. Whether such activities are essential for shorter programs, or what the ideal balance might be, remains an open question for further research.


\subsection{Impactful co-design relies on comfort and empowerment.} 
Co-design sessions thrive on open dialogue and the uninhibited sharing of ideas. When students feel at ease and empowered, they openly share their thoughts, leading to richer insights. However, with a global, diverse cohort participating remotely, there can be barriers to this communication.

\xhdr{Address any social or hierarchical barriers early on.}
Engaging with authoritative figures like mentors can sometimes make students hesitant to voice their opinions out of deference, shyness, or respect. 
To mitigate this, mentors should stress the importance of open dialogue, using affirming phrases like ``It's okay to speak your mind, we are here to guide you''. Both researchers and mentors should emphasize that all students, regardless of their background or position, are equal contributors to the co-design process. 
Mentors also addressed the challenge through open reflection in week 3 to permit an ``informal environment'' for students,
resulting in a noticeable increase in participation and also enriching our discussions. Feedback supported this approach, with one feedback from a student mentioning that ``the discussion dynamics in the group'' had improved and that it ``seems like people feel more at ease participating and sharing their thoughts and ideas.''

\xhdr{Give participants ownership over certain tasks or segments.}
Empowering participants involves more than just gathering their opinions; it means granting them control over various project aspects. In our program, we extended participants the opportunity to engage in and present background research, fostering a sense of ownership over that segment of the program. Furthermore, our approach surpassed traditional sketch-based co-design approaches; participants constructed a prototype, underwent testing, and created promotional videos for their solutions. This hands-on, project-based learning approach reinforced their responsibility and vested interest in the project.

\xhdr{Empower participants to use their strengths in their own way.}
Student diversity is not limited to backgrounds but also includes unique skills and strengths that can elevate the program. For instance, a student leveraged her significant social media presence to raise awareness and gather feedback on youth mental well-being.  Another dived into scientific research to enrich discussions with informed perspectives, while a tech-savvy participant explored futuristic research proposals, aligning them with innovative technical insights.

\xhdr{Consider what equal participation means.}
In co-design, ensuring equal participation is pivotal, yet achieving this balance is challenging. 
Participation levels among students varied across weeks and dominance by a few overshadowed others' contributions. 
This disproportionate representation can inadvertently affect the integrity of the co-design process and have detrimental effects on participants' mental well-being, potentially fostering insecurities or feelings of inadequate recognition or inclusion as well as feelings of being overwhelmed or burdened with additional activities outside of co-design sessions. 
To address this, students brainstormed potential strategies like clustering contributors based on their contribution levels to diffuse the spotlight on individual dominance or using a digital badge system for recognition.
While this idea seems promising as a means to ensure recognition and participation balance, they require future exploration.

\subsection{Co-design sessions are most impactful when participants have their goals aligned.} 
The key to effective co-design is aligning purpose and vision. When participants are united by shared objectives, they naturally exhibit stronger commitment, producing comprehensive and meaningful outcomes.

\xhdr{Consider an agile and flexible approach.}
In a rapidly evolving co-design environment, especially with a diverse set of participants, flexibility becomes paramount. For the long-term direction of our program, we opted for an agile approach, consistently realigning our trajectory based on the weekly feedback and reflections from participants. 
Our approach was agile in terms of time management and activity allocation within individual sessions. While we planned for a multitude of activities, the reality was that there was always more on the agenda than what time permitted. 
In some cases, we charged on with the planned activities that created some unresolved tension (e.g., ``I sometimes feel that we still need some time to discuss the ideas we had but we directly start brainstorming on the next step without having a real discussion.'')
In other cases, we allowed the conversations to extend (e.g., ``According to my view, time slot given for activities were less but although it was great.'')
This highlighted the complexities of managing time and resources, ensuring that the activities resonated with the participants while still maintaining the program's integrity and pacing. 
To address this, we provided students with the autonomy to vote on which activities they found most compelling or resonated with them. Such an approach ensured that sessions remained student-driven, maximizing their engagement and value derived. In the instances when time was generous, students could experience all planned activities, fully immersing in the co-design process and extracting its full benefits.

\xhdr{Provide a tailored experience to develop shared goals.}
Our co-design program was intentionally fluid in its structure, often relying on the collective input of the participants to shape its direction. From defining shared goals at the onset to taking votes on subsequent steps, participants had an active role in molding the program's trajectory. 
For example, the second week was dedicated to brainstorming and understanding everyone's goals for the program.
This responsive approach ensured the program remained in tune with participants' evolving perspectives and needs. 
However, the fluidity of the process also had its challenges as we just discussed above. 
A lesson we gleaned was the importance of striking a balance. While flexibility is vital, it becomes evident that not establishing a clearer initial direction can lead to feelings of uncertainty among participants. Therefore, while empowering participants with decision-making is beneficial, a foundational roadmap and clear explanation early on is equally crucial to guide the journey.

\subsection{Co-design could be enriched by diverse perspectives.}
The essence of co-design lies not only in collaborative efforts but also in the diverse perspectives that participants bring to the table, fusing multiple problem-solving approaches and consequently leading to innovative and holistic solutions.

\xhdr{Include activities to learn more about each other.}
Understanding and embracing diversity begins with getting to know one another. While this might not directly relate to core co-design activities, creating avenues for participants to share personal narratives and diving into the intricacies of each other's backgrounds can allow for stronger interpersonal bonds and exchanges of diverse perspectives, which is invaluable in circumventing issues like cultural insensitivity. 
We conducted activities centered on learning about each other's goals and inspirations, serving as icebreakers and setting the stage for a collaborative and cohesive work environment.
As one participant observed, ``It's eye-opening, encouraging, and insightful to know that people are concerned about the impact of technology on our mental health.'' 

\xhdr{Facilitate the exchange of diverse insights through interactive brainstorming sessions.}
Brainstorming sessions, especially in a diverse setting, become melting pots of global insights. Students, through the exchange of regional mental well-being scenarios and solutions, foster a knowledge-sharing environment. This dynamic not only allows the adaptation of ideas from one region to another but also cultivates a broader understanding of the issue at hand. As one participant aptly reflected, ``First [I] learned about Canva. [The] discussion about social issues [related to] the problem was enlightening. While discussing, I garnered a plethora of perspectives.'' This rich exchange ensures that while solutions might be rooted in local nuances, they resonate on a global scale.

\xhdr{Provide alternative means for participation.}
Given the multifaceted backgrounds of participants, it is inevitable that some may encounter technical challenges, ranging from poor internet connectivity to hardware limitations. 
To ensure inclusivity, we provided alternative participation means through recording sessions for offline viewing and maintaining a vigilant watch over the chat section. 
Leveraging online tools that are inherently designed with accessibility in mind further democratizes the process. Furthermore, archiving these sessions online not only assists those with technical difficulties but also becomes a reservoir of knowledge that students can tap into at any point.


\subsection{Co-design should be educational.} 
Co-design is more than just participation; it's a chance for comprehensive education. It seamlessly combines mentor expertise with the fresh insights of participants.

\xhdr{Provide opportunities for participants to go beyond the prepared material.}
Given the breadth of HCD and youth mental well-being, an 8-week session can only cover so much. We balanced this by focusing on essential content during live sessions and providing pre-session materials for extended learning.
Lists of background research, articles, and papers were shared, enabling students to independently explore topics at their own pace, bridge knowledge gaps, and familiarize themselves with upcoming content. This strategy provided students with the knowledge to maximize in-session discussions, while also addressing time constraints.
Unexpected insights often arose during our sessions, where facilitators injected valuable nuggets of knowledge. These range from psychological constructs (e.g., cognitive behavioral therapy) to real-world innovations targeting similar objectives (e.g., Guardian app). 
To ensure a focused progression without sidelining these insights, we often introduced these concepts briefly, and then added them to background research chat conversations for later exploration. 
This ensured the main session remained focused while still providing rich, supplementary information for the curious.

\xhdr{Include activities to learn more about research and experience research.}
Our program's inherent nature as a research study, facilitated by researchers in the field of HCD and coupled with an audience eager to set forth into the research landscape, it was crucial to introduce research-centric activities. Researcher interactions were spotlighted, giving rise to engaging and inspirational discussions such as ``Ask Me Anything'' sessions, which allowed students to candidly pose queries to both mentors and researchers, gleaning insights into career trajectories and potential opportunities. Encouraging voluntary background research also offered students a glimpse into the demands of literature reviews, providing invaluable practical experience.

\section{Conclusion}
In this study, we present our 8-week remote, co-design program where we conducted collaborative human-centered research and design aimed at guiding youth to envision and design mental well-being solutions for their communities by equipping students with the tools and methodologies to deepen their research knowledge.
Through this program, we learned that co-design with youth should be empowering through fun and engaging activities that break down social, hierarchical, and cultural barriers.
Our findings point towards the empowerment of youth in co-design that can be further enhanced through the alignment of goals, educational activities, and the youth's ownership of the design and research paths within the collaboration. 
Our lessons learned guide future co-design research work to build inclusive and accessible mental well-being technology futures for youth and by youth. 



\begin{acks}
This project would not have been possible without the students and mentors who participated in the program as well as the Microsoft Learn Student Ambassador director Pablo Veramendi and the program manager Korrie Miller.
K.P., S.S.P., and C.I. are among the student participants, G.R., D.F., E.O., and S.M.K.C., and J.S. organized and facilitated the program.
S.M.K.C. and J.S. co-mentored the students in writing this paper.
J.S. conceptualized and led the program.

\end{acks}

\bibliographystyle{ACM-Reference-Format}
\bibliography{_references}

\end{document}